# Spatial Phase and Amplitude Structuring of Beams Using a Combination of Multiple Orthogonal Spatial Functions with Complex Coefficients

Guodong Xie[1,*], Cong Liu[1], Long Li[1], Yongxiong Ren[1], Zhe Zhao[1], Yan Yan[1], Nisar Ahmed[1], Zhe Wang[1], Asher J. Willner[1], Changjing Bao[1], Yinwen Cao[1], Morteza Ziyadi[1], Ahmed Almaiman[1], Solyman Ashrafi[2], Moshe Tur[3], Alan E. Willner[1,*]

[1]Department of Electrical Engineering, University of Southern California, Los Angeles, CA 90089, USA

[2]NxGen Partners, Dallas, TX 75219, USA

[3]School of Electrical Engineering, Tel Aviv University, Ramat Aviv 69978, Israel

Corresponding email: guodongx@usc.edu, willner@usc.edu

**Introductory**: Analogous to time signals that can be composed of multiple frequency functions, we use uniquely structured orthogonal spatial modes to create different beam shapes. We tailor the spatial structure by judiciously choosing a weighted combination of multiple modal states within an orthogonal basis set, and we can tunably create beam phase and intensity "shapes" that are not otherwise readily achievable. As an example shape, we use a series of orbital-angular-momentum (OAM) functions with adjustable complex weights to create a reconfigurable spatial region of higher localized power as compared to traditional beam combining. We simulate a structured beam created by coherently combining several orthogonal OAM beams with different complex weights, and we achieve a >10X localized power density enhancement with 19 beams. Additionally, we can create unique shapes by passing a single beam through a specially designed phase and intensity mask that contains the combination of multiple OAM functions each with complex weights. Using this approach, we experimentally demonstrate a ~2.5X localized power density increase when utilizing 9 functions.

There has been much excitement about spatially structured beams, in which the phasefront and amplitude are tailored to produce specific shapes[1-3]. These shapes can usually be represented by one term of a modal basis set, which is a series of mutually orthogonal functions such as Laguerre Gaussian (LG) and Hermite Gaussian (HG)[4,5]. Additionally, a simple subset of LG modes can be orbital-angular-momentum (OAM) modes[4,6-8], which is a set of orthogonal modes that have an integer number of $2\pi$ phase changes in the azimuthal direction. The beam's phasefront twists in a helical fashion as it propagates, and the intensity profile forms a vortex-shaped ring with a central null[4,6].

These structured beams can provide us with a new tool to further create phase and intensity profiles that are not readily achievable using conventional approaches. Specifically, one can view a set of orthogonal modes each with its own complex coefficient as a series that can be combined to potentially form any arbitrary shape. This is analogous to the concept of a Fourier series[9-11] (Fig. 1a), in which any time domain signal can be constructed by adding multiple weighted orthogonal sinusoidal waves of integer frequency spacing. In our approach, we utilize a modal basis set, such as OAM, to create unique spatially structured beams by adding multiple orthogonal functions of integer OAM spacing each having a complex weighted coefficient. Such structured beams could be used in many applications for which there is a desire for tunable and unique beam intensity and/or phase shapes (e.g., medical[12], manufacturing[13], and imaging[14]).

In this paper, we propose, simulate, and experimentally demonstrate the concept of spatial phase and amplitude structuring of beams using a combination of multiple complex-weighted orthogonal functions. As a Fourier-series-like example to produce a "delta" function, we show that beams can be created with profiles exhibiting much higher localized power density than simply adding multiple fundamental Gaussian beams. We simulate the beam structuring through combining[15,16] multiple weighted OAM beams (Fig. 1b1), and we show a localized power density enhancement of >10X by combining 19 beams. We also experimentally demonstrate a proof-of-concept beam structuring by tailoring a single beam[16,17] (Fig. 1b2), and a ~2.5X localized power density increase is achieved when utilizing 9 OAM functions.

In general, beam structuring through the coherent addition of multiple beams can be represented by:

$$F(r,\theta,0) = \sum_{i=1}^{n} \alpha_i \exp(j\varphi_i) f_i(r,\theta,0), \qquad (1)$$

where $F(r,\theta,0)$ is the resultant function/beam; $f_i(r,\theta,0)$ ($i = 1,2,...,n$) is the complex field of the *i*-th contributing function/beam; $(r,\theta)$ is the cylindrical coordinate; $\alpha_i$ and $\varphi_i$ are the amplitude and temporal phase weights, respectively. The electrical field after a *z*-distance propagation, $F(r,\theta,z)$, could be derived from $F(r,\theta,0)$ according to Kirchhoff-Fresnel diffraction[18]. The goal of our structuring approach is to choose an orthogonal basis $f_i(r,\theta,0)$ and manipulate their complex coefficient $C_i = \alpha_i \exp(j\varphi_i)$, such that $F(r,\theta,z)$ has the desired spatial phase and intensity distribution. For example, to achieve a higher localized intensity, the peak of $F(r,\theta,z) \cdot F^*(r,\theta,z)$ should be maximized.

Each component function/beam in Eq. 1 can be represented at *z*=0 by:

$$f_i(r,\theta,0) = \sqrt{I(r,\theta)} \exp(j\psi(r,\theta)), \qquad (2)$$

where $I(r,\theta)$ and $\psi(r,\theta)$ are the basis's spatial intensity and phase distribution, respectively.

In order to create any desired spatial phase and intensity, we combine multiple OAM functions or beams. An OAM beam has an $\exp(j\ell\theta)$ helical transverse phase, where $\ell$ is the integer OAM order[3,5]. Thus, the spatial phase and intensity of $f_i(r,\theta,0)$ can be represented as,

$$\begin{cases} I(r,\theta) \propto r^{|\ell|} \exp(-r^2) \\ \psi(r,\theta) = \ell\theta \end{cases}. \qquad (3)$$

Our proposed approach is analogous to the concept of a Fourier series in cylindrical coordinates over the azimuthal direction, and each value of OAM represents a different phase-change rate in the azimuthal direction[9-11,19], i.e., "spatial frequency". Our simulation results show this Fourier-like time/frequency behaviour[20] in the spatial domain: (i) when combining equal-weighted multiple OAM modes, the resultant beam's azimuthal intensity distribution is a sinc-like function (Fig. 2a); (ii) comparing Fig. 2b to 2a and when the spectrum uses every other mode (i.e., sampled), the azimuthal intensity produces periodic replicas in the spatial domain; (iii) comparing Fig. 2c to 2a, a sinc-like azimuthal intensity distribution broadens when using fewer modes; (iv) a sinc-like weighted combination produces a square-like azimuthal intensity distribution (Fig. 2d).

In order to evaluate our approach, we show simulation and experimental results for one of the most basic Fourier applications. Similar to producing a "delta" function in the time domain out of multiple orthogonal frequency functions, we create spatial regions of high localized power density by adding multiple orthogonal OAM functions. We show results for two different implementations: (i) coherently combining multiple OAM beams each having a different OAM order and a potentially different complex weight, and (ii) using spatial light modulators (SLMs) to impose on a single beam a unique phase and amplitude shape that is composed of multiple orthogonal OAM functions each with a potentially different complex weight.

To analyse the localized power improvement, we use the localized power density gain (LPDG) which is defined as the peak power density ratio achieved by our approach over that achieved by simply coherent combining of fundamental Gaussian beams[21-25]. We investigate two different beam-combining scenarios: (i) all OAM beams have equal waists, and the sizes (i.e., areas) are proportional to $\sqrt{\ell + 1}$[26]; and (ii) all OAM beams have equal sizes but different waists[26,27].

Figure 3a shows the achieved LPDG using OAM orders $-\ell$ to $+\ell$ under both scenarios. The emitting aperture size is 6 mm and the transmission distance is 2.5 m. The equal size scheme shows a higher LPDG than the equal waist scheme due to more spatial overlapping. Figure 3b exhibits a lower LPDG when only 0 to $+\ell$ are used. Figure 3(c0-f5) shows that more power is concentrated in a local area as the number of combined beams increases, whereas the size only changes slightly.

Although the above simulations assume that all beams are aligned in time and therefore have the same phase coefficient $\varphi_i$, time misalignment errors may occur in practice. Figure 4a combines OAM -3 to +3 such that one beam suffers a time/phase misalignment relative to the other beams. Results show that a larger misalignment causes greater LPDG degradation, and a phase error of ~π degrades the LPDG from 5 to 2.5. Moreover, a real system may have errors in the desired amplitude coefficients $\alpha_i$, (i.e., power fluctuation) of the beams. Figure 4b combines OAM -3 to +3, such that one beam suffers a power degradation relative to the other beams. Results show that if one of the beams has zero power, the LPDG degrades from 5 to 4.2.

Since an OAM beam diverges faster as $\ell$ increases[26], the sensitivity of LPDG to OAM divergence is investigated in Figure 4c for various distances. Note that "Max OAM Order L" denotes that all OAM modes -L to +L are used. Figure 4c shows that: (i) the LPDG decreases with distance; and (ii) when L is large enough, further increasing L leads to a decrease in the LPDG. This is because larger OAM beams diverge faster, and longer distance propagation causes less overlapping between the higher-order and lower-order OAM beams. For a 50-m distance, a >10X LPDG can be achieved when 19 OAM modes are combined.

A proof-of-concept experiment is designed to demonstrate structuring of a single beam by combining multiple orthogonal functions, instead of combining multiple beams as was simulated above. In this experiment, an SLM is used to control the phase and the intensity of the incoming beam for beam shaping[17,28] (Fig. 5a) although the SLM produces power loss when imposing the functions. The emitting aperture size is 6 mm and the transmission distance is 2.5 m. Figure 5b shows the measured intensity profiles after SLM-based beam shaping, and Figures 5c and 5d show the achieved experimental and simulated LPDG, respectively. The figures show similar trends, yet an LPDG of 2.5 and 3.5 are achieved by the experiment and simulation, respectively. We believe that such a difference may be caused by the SLM's limited resolution and imperfect beam purity.

We emphasize that the proposed beam structuring is a tunable approach, as described by the following: (i) The structured beam has the spot on the edge of the entire beam. When the temporal phase weights ($\varphi_i$ in Eq. 1) of all beams are tuned simultaneously, the spot would rotate along the azimuthal direction over the beam. In the beam combining approach, these phase weights could potentially be tuned at a high speed by using an optical modulator. (ii) The LPDG could be tuned through adjusting the relative phase weights among the combined beams (Fig. 4a). (iii) The spot size of the structured beam could be tuned by varying the number of combined beams (Fig. 3c). (iv) We discuss the LPDG through observing only the peak intensity of the structured beam, and the intensity shape of the beam could also be structured by adjusting the complex weights of the combined beams/functions.

Although we use OAM as the modal basis to achieve our results, it is important to note that other orthogonal modal basis sets can in theory be used in a similar fashion. Additionally, we use OAM modes with no radial change and it is a one-dimensional complete basis over the azimuthal direction; Therefore, tunable beam structuring can be readily achieved in the azimuthal direction but not in the radial direction. A two-dimensional complete basis, such as LG modes, which considers both radial change and azimuthal change, may potentially provide an arbitrary and tunable beam structuring over a two-dimensional plane. Moreover, our approach is different than traditional beam-forming techniques since we can provide more meticulous spatial intensity and phase structuring so that the resulted beam is not necessary a fundamental Gaussian mode.

**Method**

*Beam structuring through coherent combining different OAM beams:* Previous reports have shown that, with a 'lossless' OAM mode combiner, OAM superimposing could be achieved by passing several beams at different positions through a free-space coordinate transfer system[9], thus offering the convenience of the proposed beam structuring.

*Beam structuring through phase/intensity shaping of one beam:* For the proof of concept demonstration, we achieve a structured beam by passing a regular Gaussian beam through a specially designed phase mask which can control both the intensity and the phase of the incoming beam. The specially designed phase mask is composed by three parts which can be represented by[28]:

$$\Phi(x,y)_{holo} = \left(\left(\Phi(x,y)_{phase} + \Phi(x)_{grating}\right)_{mod\ 2\pi} - \pi\right) sinc^2(1 - I(x,y)) + \pi$$

where $\Phi(x,y)_{holo}$ is the hologram loaded on the spatial light modulator (SLM); $\Phi(x,y)_{phase}$ is the designed phase mask; $\Phi(x)_{grating}$ is blazed diffraction grating in x direction; and $I(x,y)$ is the designed normalized intensity mask. After the incoming beam is reflected by the SLM, a pinhole follows to filter out the first-order diffraction of the modulated beam, which is the desired structured beam. Usually, the efficiency of the SLM is not 100%; Therefore, the blade grating helps to get rid of the unmodulated portion of the incoming beam, which stays on zero[th] order.

**References**

1. Andrews, D.L., *Structured light and its applications: An introduction to phase-structured beams and nanoscale optical forces*, Academic Press, 2011.

2. Molina-Terriza, G., Torres, J.P. & Torner, L., Twisted photons, *Nature Physics* **3**, 305-310, 2007.

3. Dickey, F.M., *Laser beam shaping: theory and techniques*, CRC Press, 2014.

4. Allen, L., Beijersbergen, M.W., Spreeuw, R.J.C. & Woerdman, J.P., Orbital angular momentum of light and the transformation of Laguerre-Gaussian laser modes, *Physical Review A* **45**, 8185, 1992.

5. Saghafi, S., Sheppard, C.J.R., & Piper, J.A., Characterising elegant and standard Hermite–Gaussian beam modes, *Optics Communications* **191**, 173-179, 2001.

6. Yao, A.M., & Padgett M.J., Orbital angular momentum: origins, behavior and applications, *Advances in Optics and Photonics* **3**, 161-204, 2011.

7. Wang, J., *et al.*, Terabit free-space data transmission employing orbital angular momentum multiplexing, *Nature Photonics* **6**, 488-496, 2012.

8. Willner, A.E., *et al.*, Optical communications using orbital angular momentum beams, *Advances in Optics and Photonics* **7**, 66-106, 2015.

9. Yao, E., *et al.*, Fourier relationship between angular position and optical orbital angular momentum, *Optics Express* **14**, 9071-9076, 2006

10. Jha, A.K., *et al.*, Fourier relationship between the angle and angular momentum of entangled photons, *Physical Review A* **78**, 043810, 2008.

11. Jack, B., Padgett, M.J. & Franke-Arnold, S., Angular diffraction, *New Journal of Physics* **10**, 103013, 2008.

12. Olesen, O.V., Paulsen, R.R., Højgaard, L., Roed, B. & Larsen, R., Motion tracking for medical imaging: a nonvisible structured light tracking approach, *IEEE Transactions on Medical Imaging* **31**, 79-87, 2012.

13. Tsai, M.J. & Hung, C.C., Development of a high-precision surface metrology system using structured light projection, *Measurement* **38**, 236-247, 2005.

14. Geng, J., Structured-light 3D surface imaging: a tutorial, *Advances in Optics and Photonics* **3**, 128-160, 2011.

15. Brignon, A., *Coherent Laser Beam Combining*, Wiley-VCH, 2013.

16. Motamed, M. & Runborg, O., Taylor expansion and discretization errors in Gaussian beam superposition, *Wave Motion* **47**, 421-439, 2010.

17. Stilgoe, A.B., Kashchuk, A.V., Preece, D. & Rubinsztein-Dunlop, H., An interpretation and guide to single-pass beam shaping methods using SLMs and DMDs, *Journal of Optics* **18**, 065609, 2016.

18. Kirchhoff, G., Zur theorie der lichtstrahlen, *Annalen der Physik* **254**, 663-695, 1883.

19. Jack, B., *et al.*, Demonstration of the angular uncertainty principle for single photons, *Journal of Optics* **13**, 064017, 2011.

20. Bracewell, R., *The fourier transform and iis applications*, New York, 1965.

21. Kozlov, V.A., Hernandez-Cordero, J., & Morse, T.F., All-fiber coherent beam combining of fiber lasers, *Optics Letters* **24**, 1814-1816, 1999.

22. Cheo, P.K., Liu, A., & King, G.G., A high-brightness laser beam from a phase-locked multicore Yb-Doped fiber laser array, *IEEE Photonics Technology Letters* **13**, 439-441, 2001

23. Sabourdy, D., *et al.*, Efficient coherent combining of widely tunable fiber lasers, *Optics*

12. Olesen, O.V., Paulsen, R.R., Højgaard, L., Roed, B. & Larsen, R., Motion tracking for medical imaging: a nonvisible structured light tracking approach, *IEEE Transactions on Medical Imaging* **31**, 79-87, 2012.

13. Tsai, M.J. & Hung, C.C., Development of a high-precision surface metrology system using structured light projection, *Measurement* **38**, 236-247, 2005.

14. Geng, J., Structured-light 3D surface imaging: a tutorial, *Advances in Optics and Photonics* **3**, 128-160, 2011.

15. Brignon, A., *Coherent Laser Beam Combining*, Wiley-VCH, 2013.

16. Motamed, M. & Runborg, O., Taylor expansion and discretization errors in Gaussian beam superposition, *Wave Motion* **47**, 421-439, 2010.

17. Stilgoe, A.B., Kashchuk, A.V., Preece, D. & Rubinsztein-Dunlop, H., An interpretation and guide to single-pass beam shaping methods using SLMs and DMDs, *Journal of Optics* **18**, 065609, 2016.

18. Kirchhoff, G., Zur theorie der lichtstrahlen, *Annalen der Physik* **254**, 663-695, 1883.

19. Jack, B., *et al.*, Demonstration of the angular uncertainty principle for single photons, *Journal of Optics* **13**, 064017, 2011.

20. Bracewell, R., *The fourier transform and iis applications*, New York, 1965.

21. Kozlov, V.A., Hernandez-Cordero, J., & Morse, T.F., All-fiber coherent beam combining of fiber lasers, *Optics Letters* **24**, 1814-1816, 1999.

22. Cheo, P.K., Liu, A., & King, G.G., A high-brightness laser beam from a phase-locked multicore Yb-Doped fiber laser array, *IEEE Photonics Technology Letters* **13**, 439-441, 2001

23. Sabourdy, D., *et al.*, Efficient coherent combining of widely tunable fiber lasers, *Optics*

*Express* **11**, 87-97, 2003.

24. Fan, T.Y., Laser beam combining for high-power, high-radiance sources, *IEEE Journal of Selected Topics in Quantum Electronics* **11**, 567-577, 2005.

25. Sprangle, P., Peñano, J., Hafizi, B., & Ting, A., Incoherent combining of high-power fiber lasers for long-range directed energy applications, *Journal of Directed Energy* **2**, 273-284, 2007.

26. Phillips, R.L. & Andrews, L.C., Spot size and divergence for Laguerre Gaussian beams of any order, *Applied Optics* **22**, 643-644, 1983.

27. Born, M. & Wolf, E., *Principles of optics: electromagnetic theory of propagation, interference and diffraction of light*, CUP Archive, 2000.

28. Leach, J., Dennis, M.R., Courtial, J. & Padgett, M.J., Vortex knots in light, *New Journal of Physics* **7**, 55, 2005.

**Acknowledgements**

This work is supported by National Science Foundation (NSF).

**Author contributions**

The authors declare no competing financial interests. All the authors were involved in the data analysis, and all contributed to writing of the article. G.X. and C.L. performed the measurement of the beam shaping. G.X., L.L., Y.R., Z.Z., and A.E.W. performed the scheme design. The project was supervised by M.T. and A.E.W.

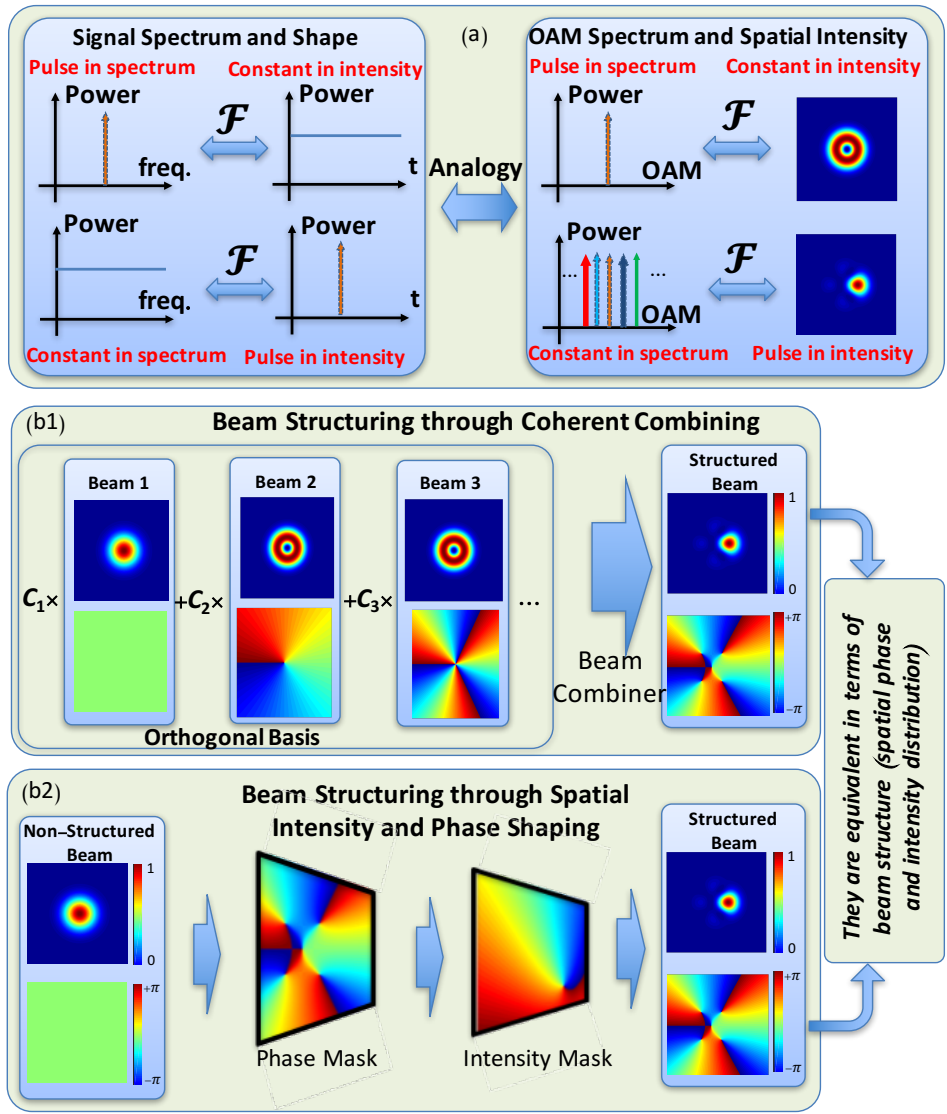

**Figure 1 | Principle and Approaches for Beam Structuring**. (a) Analogy between the "frequency-time" relationship of a signal and the "OAM spectrum-spatial intensity" relationship of a beam. (b1) Beam structuring through coherently combining several beams from an orthogonal modal basis with complex coefficients $C_i$. (b2) Beam structuring through phase and intensity shaping of a single beam.

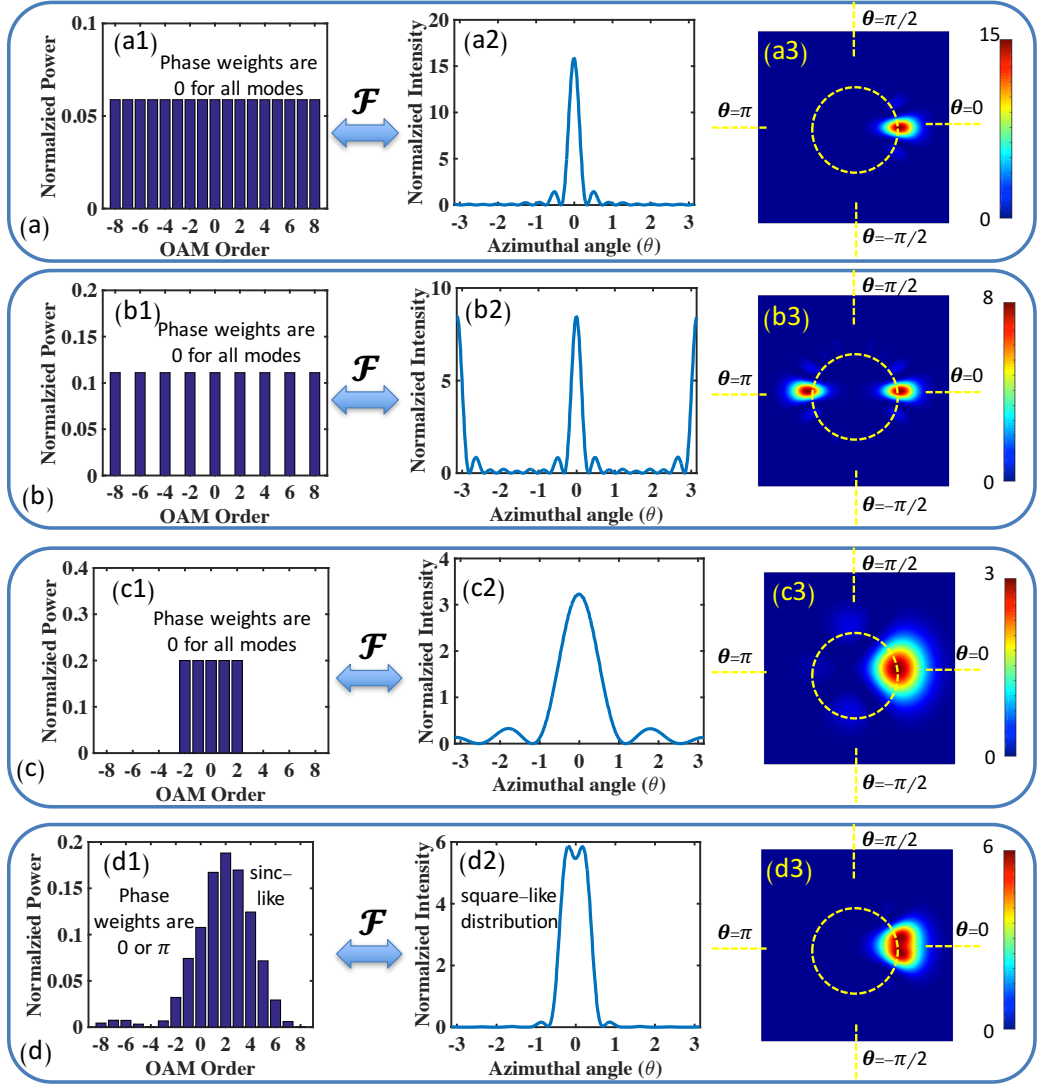

**Figure 2 | Simulation Results of Fourier Relationship Examples Between the OAM Spectrum and Intensity Distribution of the Structured Beam in the Azimuthal Direction.** (a1, b1, c1, d1) OAM spectrum; The temporal phase weights for all OAM modes in a1, b1, c1 are 0; The temporal phase weights in d1 are 0 for OAM -3 to +7 and $\pi$ for the other modes. (a2, b2, c2, d2) Normalized intensity along the azimuthal direction over the beam. (a3, b3, c3, d3) Intensity profiles of the structured beam. The yellow rings indicate the radial position where (a2, b2, c2, d2) are measured.

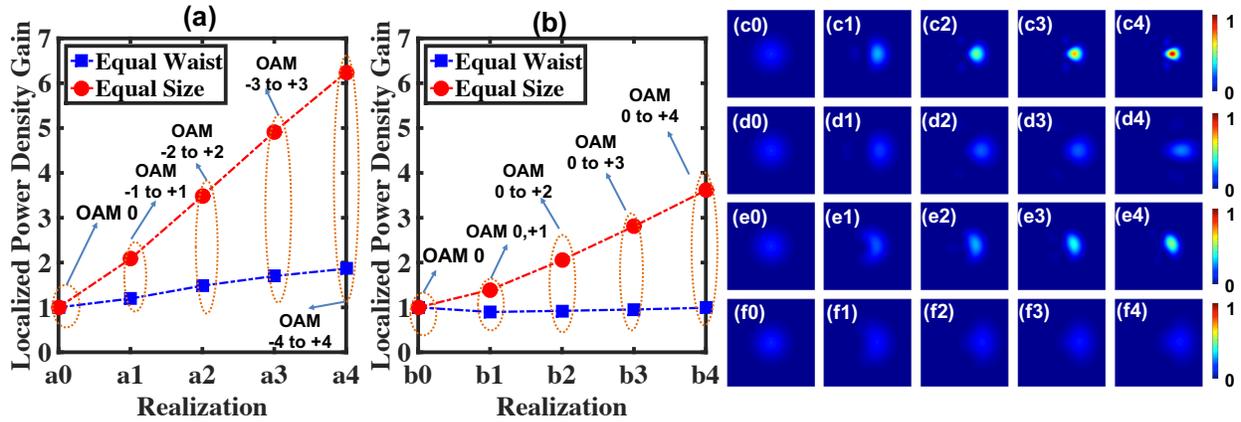

**Figure 3 | Simulated Localized Power Density Gain Achieved by Beam Structuring.** (a) Combination of OAM $-\ell$ to $+\ell$ beams. (b) Combination of OAM 0 to $+\ell$ beams. Intensity profiles of the structured beams: (c0-c4) Combining of OAM $-\ell$ to $+\ell$ beams with the same beam size; (d0-d4) Combining of OAM $-\ell$ to $+\ell$ beams with the same beam waist; (e0-e4) Combining of OAM 0 to $+\ell$ beams with the same beam size; and (f0-f4) Combining of OAM 0 to $+\ell$ beams with the same beam waist. All OAM beams have the same complex weight. Emitting aperture size: 6 cm; transmission distance: 100 m.

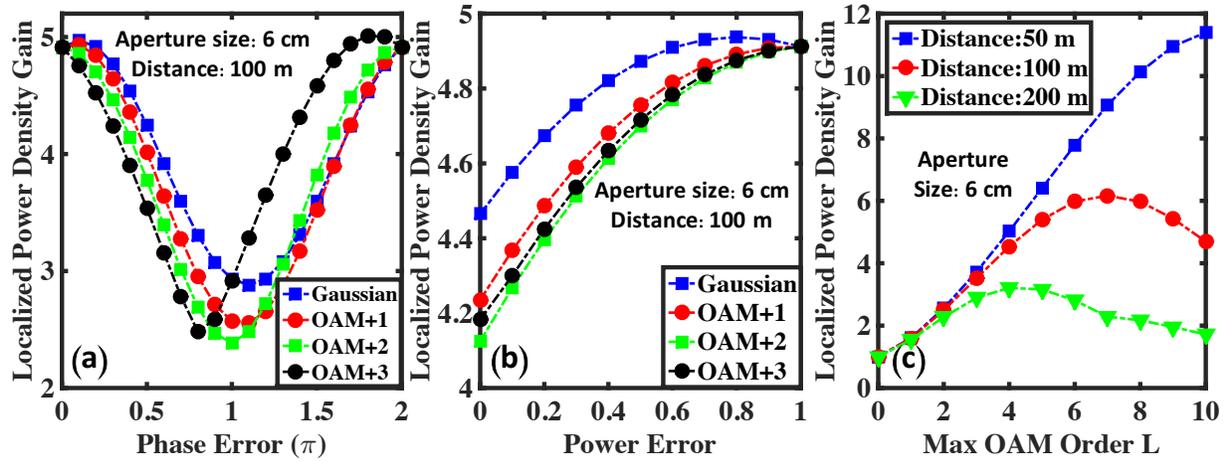

**Figure 4 | Simulated Localized Power Density Gain (LPDG) Performance Under Different Scenarios**. (a) LPDG degradation due to the time/phase misalignment of one of the modes when OAM -3 to +3 are combined. (b) LPDG degradation due to the power fluctuation of one of the modes when OAM -3 to +3 are combined. (c) LPDG as a function of the maximum OAM orders (L) for beam structuring at various distances. "Maximum OAM Orders L" denotes that OAM -L to +L are used for beam combining. For a fair comparison, the total power for all tests is the same.

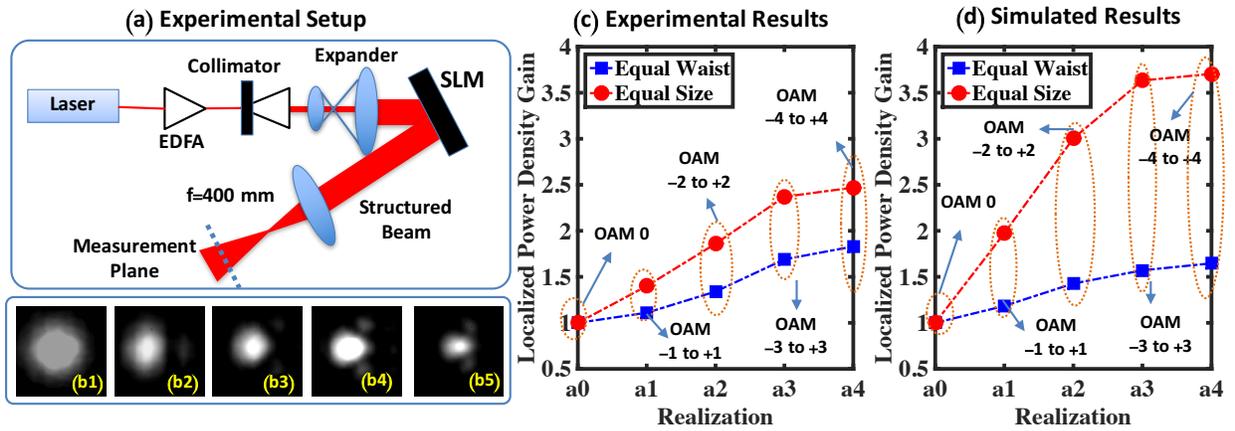

**Figure 5. Experimental Beam Structuring for Achieving Localized Power Density Gain (LPDG).** (a) Experimental setup. EDFA: Erbium-doped fibre amplifier; SLM: spatial light modulator. (b1-b5) Experimental intensity profiles of the structured beam using superposition function of: OAM 0, OAM -1 to +1, OAM -2 to +2, OAM -3 to +3, and OAM -4 to +4, respectively. Comparison between the experimental (c) and simulated (d) LPDG achieved through beam structuring. Emitting aperture size: 6 mm; transmission distance: 2.5 m.